# Local Manipulation and Topological Phase Transitions of Polar Skyrmions

Linming Zhou[1], Yongjun Wu[*, 1, 2], Sujit Das[3], Yunlong Tang[3], Cheng Li[1], Yuhui Huang[*, 1], He Tian[4], Long-Qing Chen[5], Ramamoorthy Ramesh[*, 3], Zijian Hong[*, 1, 2]

[1] Lab of Dielectric Materials, School of Materials Science and Engineering, Zhejiang University, Hangzhou, Zhejiang 310027, China

[2] Cyrus Tang Center for Sensor Materials and Applications, State Key Laboratory of Silicon Materials, Zhejiang University, Hangzhou, Zhejiang 310027, China

[3] Department of Materials Science and Engineering, University of California, Berkeley, CA 94720, USA

[4] Center of Electron Microscopy, State Key Laboratory of Silicon Materials, School of Materials Science and Engineering, Zhejiang University, Hangzhou, Zhejiang 310027, China

[5] Department of Materials Science of Engineering, The Pennsylvania State University, University Park, PA 16802, USA

Topological phases such as polar skyrmions have been a fertile playground for ferroelectric oxide superlattices, with exotic physical phenomena such as negative capacitance. Herein, using phase-field simulations, we demonstrate the local control of the skyrmion phase with electric potential applied through a top electrode. Under a relatively small electric potential, the skyrmions underneath the electrode can be erased and recovered reversibly. A topologically protected transition from the symmetric to asymmetric skyrmion bubbles is observed at the edge of the electrode. While a topological transition to a labyrinthine domain requires a high applied potential, it can switch back to the skyrmion state with a relatively small electric potential. The topological transition from +1 to 0 occurs before the full destruction of the bubble state. It is shown that the shrinking and bursting of the skyrmions leads to a large reduction in the dielectric permittivity, the magnitude of which depends on the size of the electrode.

* Corresponding authors: Y.W. (yongjunwu@zju.edu.cn), Y. H. (huangyuhui@zju.edu.cn) R.R.(rramesh@berkeley.edu), Z.H. (hongzijian100@zju.edu.cn)

# Introduction

Complex ferroelectric topological patterns have garnered enormous interest recently, driven by the formation of polar vortices [1-5], flux-closure domains [6, 7], polar spirals [8, 9], skyrmions [10-12], and merons [13], etc., which have largely rejuvenated the understanding of ferroelectrics and the associated theoretical frameworks [14]. These polar textures exhibit exotic physical phenomena such as negative capacitance [15-17], chirality [18-22], ultrafast light-induced supercrystal formation [23], etc. In particular, the polar skyrmion, a nontrivial whirl-like structure with a topological charge of ±1, has been discovered in the $PbTiO_3/SrTiO_3$ (PTO/STO) superlattice system. It has triggered considerable attention for potential nano-electronic device applications, like in the non-centrosymmetric ferromagnetic counterparts [24, 25]. Previous studies have shown that the polar skyrmion in the PTO/STO system is topologically protected during electrical switching, akin to those observed in the ferromagnetic systems [16]. The dynamics of the skyrmion switching process could also induce transitions from a skyrmion to other topological states of matter in a reversible fashion [11]. Despite all these exciting discoveries, one intriguing question has yet to be answered: how can we locally move/manipulate the polar skyrmion? This could stimulate the next level of interest in technological applications such as ferroelectric racetrack memories or other electronic devices.

# Results

In this article, we report the local electric erasure/recovery of polar skyrmions via an external electric field through a narrow top electrode contact. Details of the phase-field simulations are given in the Methods. The simulation setup is described in **Fig. 1**, where the $PbTiO_3/SrTiO_3$ (PTO/STO) superlattice is epitaxially grown on top of a (001)-STO substrate. A three-dimensional mesh of 320×320×350 is built, with each grid representing 1 unit cell. A bottom electrode is introduced between the film and substrate, while a narrow electrode with designed width ($d_0$= 8 nm and 32 nm) along the *X*-dimension is deposited on top of the film. The top view shows the equilibrium polar skyrmion structure in the PTO layer (**Fig. 1b**), which is consistent with previous reports [10, 16]. **Fig. 1(c)** shows the magnified view of the in-plane vector mapping, which indicates the formation of a hedgehog-like skyrmion structure. To quantify the topological feature of the polar skyrmion, the Pontryagin density $q$ is calculated, i.e., $q = \frac{1}{4\pi}\vec{P}\cdot(\frac{\partial \vec{P}}{\partial x} \times \frac{\partial \vec{P}}{\partial y})$, where $\vec{P}$ is the polarization vector. The topological charge can be obtained by the surface integral of the Pontryagin density over *xy* space, $Q = \iint q dx dy$ [12, 27-28]. The planar view of the Pontryagin density for the top PTO layer is plotted in **Fig. 1(d)**. It shows the formation of ring-like structures, which are planar projections of the Neel component of the dipolar distribution of the Pontryagin density for smooth skyrmion bubbles [11]. While the line profile shows two peaks on the two edges of the ring (**Fig. 1e**), for each ring, the total topological charge is +1. An electric potential is then applied through the top electrode, with the potential profile given in **Fig. 1(f)**.

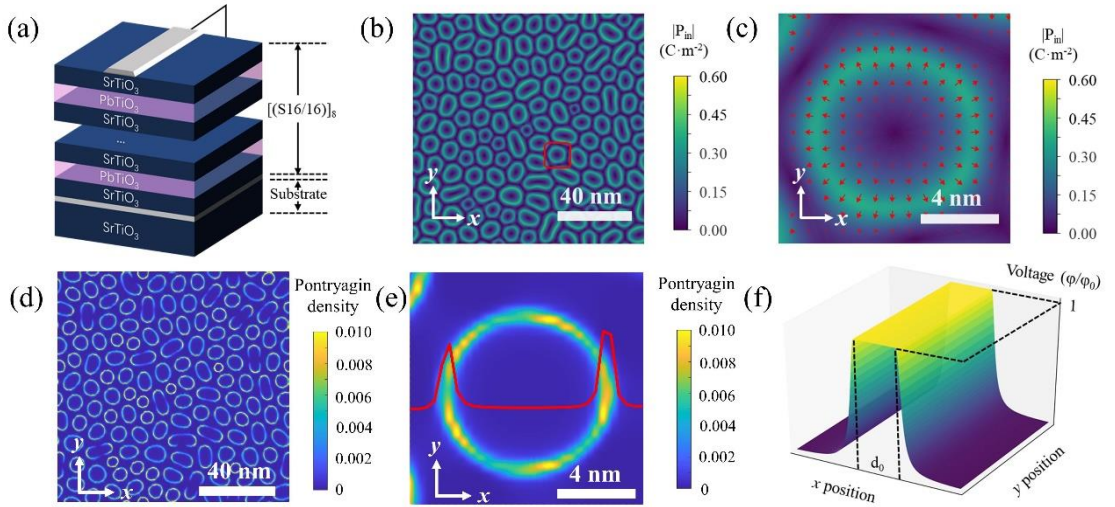

**Fig. 1 Initial setup, topological feature and the applied potential profile.** (**a**) Schematics of the PTO/STO superlattice system. (**b**) Planar view of the in-plane polarization magnitude from phase-field, showing the formation of skyrmion bubbles. (**c**) Magnified view of the polar skyrmion bubble overlaid with the in-plane polar vector. (**d**) Planar view of the Pontryagin density distribution. (**e**) Magnified view of the Pontryagin density

distribution on a single bubble, with line profile overlaid. **(f)** The applied voltage profile, the width of the electrode is $d_0$, while the magnitude of the voltage is $\varphi_0$.

In the first study, a +6 V is applied through an 8 nm wide top electrode. The kinetic evolution pathway of the skyrmions is shown in **Fig. 2(a)-2(c)**. The skyrmions underneath the electrode region shrink after 100 timesteps, and then gradually disappear. In the neighboring regions, the skyrmions are pushed away from the electrode, which become asymmetric where it is darker near the electrode side and brighter on the other side. Eventually, all the skyrmions underneath the electrode region are erased. The dynamics of the evolution pathway is presented in **supplementary video S1**. The polar structure of the asymmetric skyrmion in the vicinity of the electrode is highlighted in **Fig. 2(d)**; a large in-plane polarization is observed which points away from the planar electrode. While close to the electrode side, the in-plane polarization becomes much smaller in magnitude. To understand this phenomenon, the electric field distribution of the top PTO layer is plotted (**Fig. S2a and S2b**), indicating that the electric field is predominantly pointing downwards underneath the electrode region, which triggers the switching of skyrmions to form a simple $c^-$ domain. In the neighboring regions, a large in-plane electric field can be observed due to the sharp potential transition from +6 V to 0 V, which drives the asymmetric skyrmion transition and pushes them away from the electrode. The topological feature of the asymmetric bubble is given in **Fig. 2(e)**. The distribution of the Pontryagin density is asymmetric with the left side of the ring having a higher Pontryagin density, corresponding to a single peak on the left from the line plot. The topological charge for the asymmetric bubble is also +1, demonstrating that the symmetric to asymmetric bubble transition is also topologically protected. The Pontryagin density for the asymmetric bubble is further plotted with different switching timesteps (**Fig. 2f**). It can be observed that a single peak formed after 200 timesteps, which shows a right shift with continuous evolution of the asymmetric skyrmion. Meanwhile, as a comparison, for the skyrmion bubbles underneath the electrode, the Pontryagin density is symmetric on the two sides of the ring, which increases with the decreasing of the bubble size until it bursts after ~$10^4$ timesteps (**Fig. S2d**).

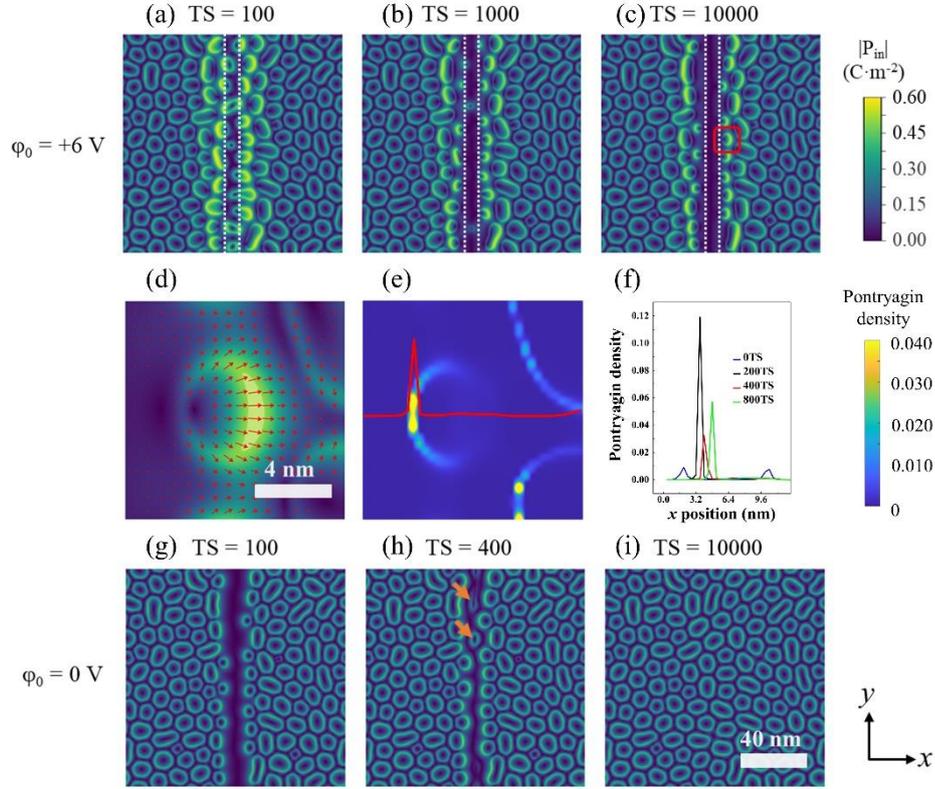

**Fig. 2 Kinetics and topological transition of the skyrmion switching and recovery under a small applied potential of 6 V and narrow electrode of 8 nm.** (a)-(c) The kinetic evolution pathway after $10^2$, $10^3$ and $10^4$ timesteps, respectively. The white dashed lines mark the electrode region. (d) Magnified view of the asymmetric skyrmion overlaid with the inplane polar vector. (e) Pontryagin density distribution on an asymmetric bubble, with line profile overlaid. (f) Line plots of the Pontryagin density over different timesteps. (g)-(i) The relaxation kinetics after the field is removed, after 100, 400 and $10^4$ timesteps, respectively. The arrows indicate the nucleation of new skyrmions.

Upon removal of the applied electric field, the asymmetric skyrmions first expand and relax to symmetric skyrmions, then the nucleation and growth of new skyrmions can be observed to fill in the regions underneath the original electrode (**Fig. 2g-2i**). After $10^4$ timesteps, the system reverts to a fully skyrmionic state. Thus, we have demonstrated the ability to locally erase and reset the skyrmion state, which is electrically controllable and reversible.

On the other hand, when the applied bias is large (e.g., +9 V, **Fig. 3**), the skyrmion shrinking and dissipation process is also observed underneath the electrode region, similar to the previous case. However, in the neighboring regions, the skyrmions "melt" and merge to form long stripe domains along the $Y$ direction. After $10^4$ timesteps, all the skyrmions underneath the electrode regions are erased, while the neighboring regions are filled with long stripes (**Fig. 3c**). These stripe domains have a large unidirectional in-plane polarization pointing away from the electrode region (**Fig. S3a** and **S3b**). This can be rationalized since under +9 V, the in-plane field is much higher as compared to the +6 V case, which is sufficient to switch all the skyrmions to a unidirectional in-plane polarization and connect them (**Fig. S3c** and **S3d**). The topological feature after 100 timesteps is plotted in **Fig. 3(d)**. It can be seen that the Pontryagin density of the skyrmions underneath the electrode behaves similar to the previous case which increases in magnitude with decreasing the bubble size. Meanwhile, an interesting topological pattern is observed in the vicinity of the electrode where the rings show alternating positive and negative Pontryagin density on the two sides **(Fig. 3e)**, which gives rise to a net topological charge of 0. This demonstrates the switching of the topological pattern even before the full destruction of the bubble structure. It can also be observed that the shape of the Pontryagin density is preserved during the consequent switching process, while the peak value shrinks and completely disappears when the bubbles join to form long stripes **(Fig. 3f)**.

When the applied potential is removed, the stripe domains expand towards the electrode region, forming labyrinthine domains to reduce the depolarization field in this system. Unlike the unidirectional stripes formed in the initial switching process, these labyrinthine domains have alternating in-plane polarization components to minimize the depolarization field. New stripes nucleate and grow underneath the electrode region, from the vicinity of the old stripes to the center of the electrode region (**Fig. 3g** and **3h**). The detailed switching and recovery process is further provided in **supplementary videos S2**. Previously, topological transitions from vortex to skyrmions by thermal or electrical driving forces have been widely reported [11, 30, 31]. Here in this study, we have demonstrated an electric field-driven localized topological transition from skyrmions to **labyrinthine domains**, with both states being stable at room temperature and zero field. The net skyrmion number of the whole simulation system (which is calculated by integrating the Pontryagin density for the entire top PTO surface) is shown for different switching processes (**Fig. 3i**). It can be observed that with a small applied voltage bias and a narrow electrode, the total skyrmion number decreases first as the skyrmion bubbles burst but recovers after the applied bias is removed. Under a high electric field with a wide electrode, the total skyrmion number drops significantly which cannot be restored even after the applied electric field is removed.

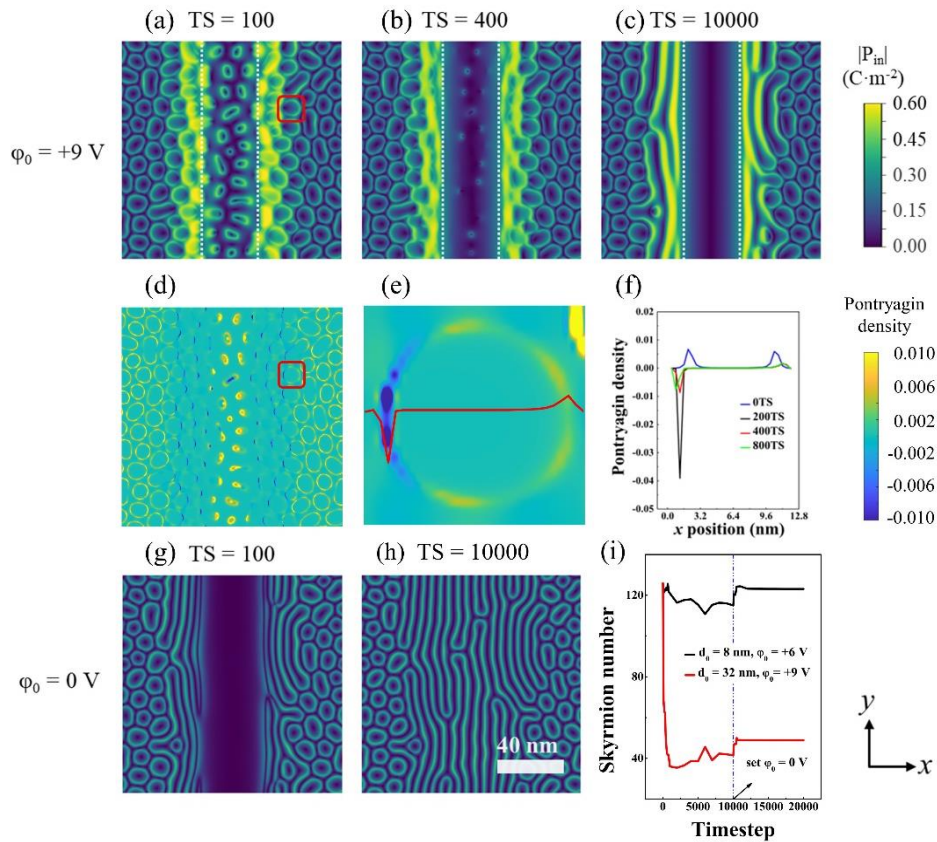

**Fig. 3 Kinetics and topological transition of the skyrmion switching and recovery under high potential of +9 V and wide electrode of 32 nm.** (a)-(c) The kinetic evolution after 100, 400 and $10^4$ timesteps. The white dashed region is the electrode. (d) 2-D plot of the Pontryagin density on the top PTO layer after 100 timesteps. (e) Magnified view of the Pontryagin density distribution on a bubble, with line profile overlaid, showing a distinct

topological feature with negative Pontryagin density on the left and positive Pontryagin density on the right. **(f)** Evolution of the line plot of the Pontryagin density. **(g)**-**(h)** The relaxation kinetics after the field is removed, after 100 and $10^4$ timesteps. The formation of stable long stripes is observed. **(i)** Comparison of the evolution of the total skyrmion number during the switching and recovery process under low and high bias.

A cycling test is further performed (**Fig. 4**) to showcase the controllable reverse transition, starting from the state after poling by +9 V and relaxed to zero after $10^4$ timesteps (**Fig. 4a**). An opposite bias of -6 V is applied later (**Fig. 4b**), which melts the vortex and labyrinthine domains to skyrmion bubbles through a Rayleigh-Plateau mechanism (see supplementary video **S3** for details), as has been reported previously [11]. After the field is removed, these residual small bubbles expand to become normal skyrmions, while the nucleation of skyrmions in other regions is also observed (**Fig. 4c**). On the two sides near the electrode, the labyrinthine domains remain since the in-plane field generated under -6 V is not large enough to erase them. Then, in the second cycle, +9 V is applied again to erase the skyrmion bubbles (**Fig. 4d**) with pure *c*-domain underneath the electrode region, which switch back to the labyrinthine state again after the field is removed (**Fig. 4e**). A potential of -6 V is applied to start the second cycle, leading to the topological transition between labyrinthine domains and skyrmions (**Fig. 4f**) and the skyrmion state is stable when the applied field is removed (**Fig. 4g**). Thus, we have realized the localized reversible transitions between skyrmions and labyrinthine domains through control of the magnitude of the applied voltage.

To further understand the properties accompanying the topological transitions, the dielectric properties underneath the electrode regions of the different states are further plotted in **Fig. 4(h)**. The initial dielectric constant of the skyrmion bubble is ~650, which agrees with the previous experimental measurements and theoretical calculations [16]. It can be observed that with the increase of the applied electric field, the dielectric permittivity decreases. This is largely due to the shrinking and switching of the skyrmions, which decreases the area with negative permittivity inside the PTO layers, consistent with previous studies [13, 14]. Notably, under the same applied voltage, generally the wider the electrode, the larger the dielectric permittivity. This can be understood since under the same, nominal out-of-plane electric field, the remaining skyrmion density/area is much larger for the case with a wider electrode than a narrower electrode, as evidenced by **Fig. S4(a)** and **S4(b)**. This will contribute to extra, negative capacitance regions that will ultimately increase the overall dielectric property. When the electrode is thin (e.g., 8 nm), with a large nominal applied field (i.e., +800 kV/cm), the dielectric constant underneath the electrode shows an 80% reduction due to the erasure of the polar skyrmions with large negative capacitance regions, indicating the large tunability of the device.

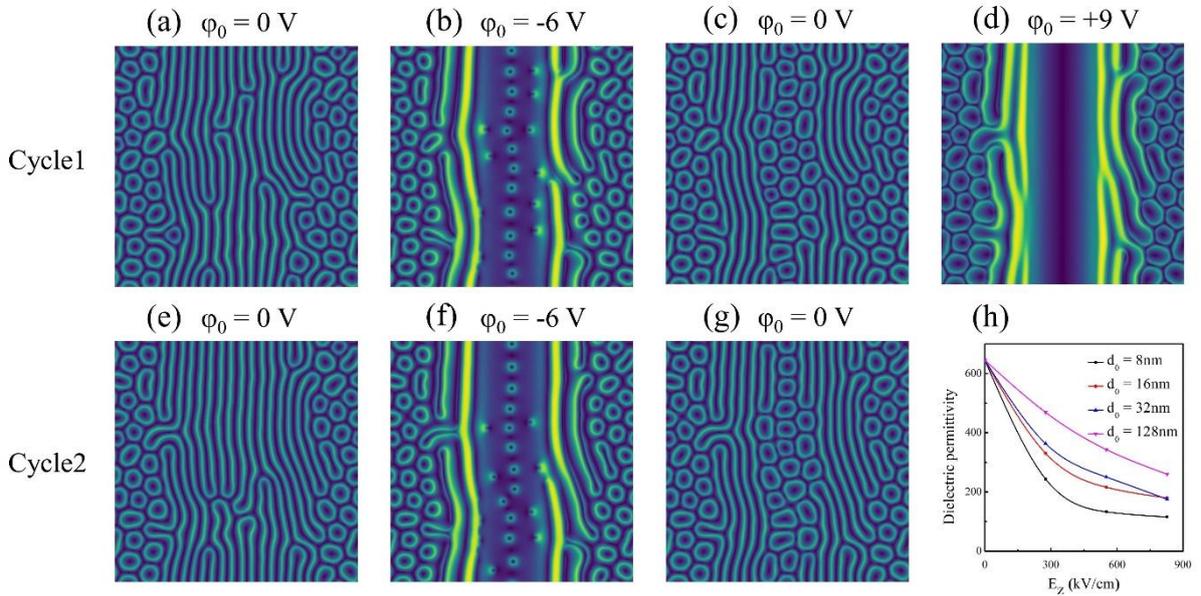

**Fig. 4 Cycling test and reversible transition between labyrinthine state and skyrmion.** **(a)**-**(d)** The kinetic evolution of the first cycle, after applying -6 V to erase the labyrinthine domains to form stable skyrmion when field is removed, while applying +9 V could erase the skyrmions and form labyrinthine state. **(e)**-**(g)** The second cycle from vortex to skyrmion after applying -6 V. **(h)** local dielectric permittivity as a function of nominal applied electric field through a narrow top electrode.

**Discussion**

In conclusion, we report the local control of polar skyrmions through a top electrode in the PTO/STO superlattice system. Under a small applied potential with a narrow electrode geometry, the skyrmions underneath the electrode can be locally erased while the skyrmions in the neighboring regions are repelled by the electrode, giving rise to asymmetric shapes with in-plane polarization mainly pointing away from the electrode, both the symmetric and asymmetric skyrmion bubbles exhibit a topological charge of +1, showing that the asymmetric to symmetric skyrmion transition is topologically protected. The skyrmion state is recovered when the potential is removed. Meanwhile, with a large applied potential and a wide electrode, labyrinthine domains form on the two sides of the electrode, and remain stable even when the applied potential is removed. Interestingly, in this case, the topological transition from +1 to 0 occurs even before the full destruction of the bubble state, with a negative/positive Pontryagin density on the left/right side of the bubble. As a comparison, the total topological charge of the system can be recovered under small bias, while under large applied potential and wide electrode, the total topological charge is largely decreased which can't be restored when the field is removed. The labyrinthine states can be switched back to the skyrmion state after applying a relatively smaller potential. Thus, we realized a reversible transition between skyrmion and labyrinthine states through controlling the distribution of the applied potential. The dielectric permittivity is further calculated, which shows a large reduction/tunability under a high field. This can be attributed to the erasure of the skyrmion regions with negative permittivity. We envision this work to spur further interests in the skyrmion physics in ferroelectric systems as well as the potential applications towards nano-electronic device applications.

**Methods:**

Phase-field simulations.

Phase-field simulations are performed for the [(PbTiO$_3$)$_{16}$/(SrTiO$_3$)$_{16}$] superlattices on SrTiO$_3$ (001) substrate, by solving the time dependent Ginzburg-Landau equations [31-33]:

$$\frac{d\vec{P}}{dt} = -L \frac{\delta F(\vec{P}, \nabla \vec{P})}{\delta \vec{P}} \tag{1}$$

Where $\vec{P}$ is the spontaneous polarization vector, $t$ is the evolution timestep. The kinetic coefficient $L$ is related to the domain wall mobility. The free energy $F$ can be expressed by the volume integral of the elastic, electric, Landau and gradient energy densities, i.e.,

$$F(\vec{P}, \nabla \vec{P}) = \int (f_{elas} + f_{elec} + f_{Land} + f_{grad}) dV \tag{2}$$

The Landau chemical energy density can be written as:

$$f_{Land} = \alpha_{ij} P_i P_j + \beta_{ijkl} P_i P_j P_k P_l + \cdots \tag{3}$$

In this study, the Landau polynomial is expanded to the sixth order, with the coefficients taken from previous literatures [26, 34].

The electric energy density is calculated by:

$$f_{elec} = -\frac{1}{2} E_i P_i + k_{ij} E_i E_j \tag{4}$$

Where $k_{ij}$ is the background dielectric constant, which is set as 40 in this study [35-37]. The local electric field $E$ can be obtained by: $E_i = -\nabla_i \varphi$. The closed boundary conditions are set such that the electric potential $\varphi$ is zero and the applied potential on the film/substrate interface and the top of the thin film, respectively.

The elastic energy density can be obtained:

$$f_{elas} = C_{ijkl}(\varepsilon_{ij} - \varepsilon_{ij}^0)(\varepsilon_{kl} - \varepsilon_{kl}^0) \tag{5}$$

Where $C_{ijkl}$ is the elastic stiffness tensor, $\varepsilon_{ij}^0$ is the eigen strain due to the ferroelectric phase transition, $\varepsilon_{ij}^0 = Q_{ijkl}P_k P_l$, $Q$ is the electrostriction coefficient tensor. The strain tensor $\varepsilon_{ij}$ can be calculated by solving the elastic equilibrium equation. An iterative perturbation method is used to consider the elastic anisotropy for the PTO and STO layers [38]. The pseudocubic lattice constants for PTO and STO are set as 3.957 Å and 3.905 Å, respectively to determine the lattice mismatch. Thin film elastic boundary condition is used where the out-of-plane stress on the top of the thin film is set as zero while the displacement on the bottom of the substrate far away from the electrode/thin film interface is zero.

The gradient energy density is further expressed as:

$$f_{grad} = g_{ijkl}\nabla_j P_i \cdot \nabla_l P_k \tag{6}$$

Where $g$ is the gradient coefficient tensor.

Detailed numerical calculations and the simulation parameters can be found in the published literature [8, 11, 26]. The simulation system is discretized into a 3-D mesh of 320×320×350, with each grid representing 1 unit cell. A periodic boundary condition is applied on the in-plane dimensions, while a superposition method is used on the out-of-plane direction. The out-of-plane direction consists of 30 grids of substrate, 272 grids of thin films with periodic stacking of PTO$_{16}$/STO$_{16}$ and 48 layers of air. The normalized timestep is 0.01 in this study.

The local dielectric constant is obtained by: $\varepsilon_{33} = \frac{\Delta P_3}{\varepsilon_0 \Delta E_3} + \varepsilon_b$, while the macroscopic dielectric constant can be calculated by $\overline{\varepsilon_{33}} = \frac{<\Delta P_3>d}{\varepsilon_0 \varphi} + \varepsilon_b$, where $<\Delta P_3>$ is the average of the change in out-of-plane polarization underneath the electrode, while $d$ is the thickness of the film, and $\varphi$ is the applied electric potential.

Sample preparation.

[(PbTiO$_3$)$_n$/(SrTiO$_3$)$_n$]$_m$ superlattices were synthesized on TiO$_2$-terminated single-crystalline SrTiO$_3$ (001) substrates via reflection high-energy electron diffraction (RHEED)-assisted pulsed-laser deposition (KrF laser). The growth temperature and oxygen pressure for the bottom SrTiO$_3$ layer were 700 ℃ and 50 mTorr, respectively. The PbTiO$_3$ and the top SrTiO$_3$ were grown at 610 ℃ in 100 mTorr oxygen pressure. For all materials, the laser fluence was 1.5 J/cm$^2$ with a repetition rate of 10 Hz. RHEED was used during the deposition to ensure the maintenance of a layer-by-layer growth mode for both the PbTiO$_3$ and SrTiO$_3$. The specular RHEED spot was used to monitor the RHEED oscillations. After deposition, the heterostructures were annealed for 10 minutes in 50 Torr oxygen pressure to promote full oxidation and then cooled down to room temperature at that oxygen pressure.

Scanning transmission electron microscopy.

Plan-view samples of the SrTiO$_3$/PbTiO$_3$ superlattices for STEM experiments were prepared by gluing a 2.5 mm ×2.5 mm film on a 3 mm molybdenum grid. Cross-sectional samples of the SrTiO$_3$/PbTiO$_3$ superlattices for TEM experiments were prepared by gluing two 3 mm ×2 mm films face-to-face on a 3 mm molybdenum grid. Samples were then ground, dimpled, and, finally, ion milled. A Gatan PIPS II was used for the final ion milling. Before ion milling, the samples were dimpled down to less than 20 μm. HAADF-STEM images were recorded by using the Cs-corrected TEAM1 FEI Titan microscope working at 300 kV at room temperature. A HAADF detector resulting in "Z-contrast" images were used to record the HAADF-STEM images. The beam convergence angle was 17 mrad. The dark-field diffraction contrast image was recorded using a TitanX microscope (FEI) working at 300 kV.

**Acknowledgements**

A start-up grant from Zhejiang University is acknowledged (Z. H.). This work was financially supported by the Fundamental Research Funds for the Central Universities (No. 2021FZZX003-02-03). The phase-field simulation is performed on Shanghai Supercomputing Center (SSC), specifically on the MoFang III cluster.

**Author contributions**

Z. H. conceived the idea. L. Z. performed the phase-field simulations with the help from L. C. S. D. carried out the synthesis of the superlattice samples. Y. T. and H. T. performed the TEM characterization. L. Z., Y. W., S. D., Y. T., L. -Q. C., R. R., Z. H. analyzed the data and co-wrote the manuscript. Y. W., Y. H., L. -Q. C., R. R., and Z. H. supervised the research. All authors contributed to the discussions and manuscript preparations.

**Ethics declarations**

Competing interests

The authors declare no competing interests.